\documentclass{iopconfser}

\usepackage{amsmath,amssymb}
\usepackage{graphicx}

\newcommand{\gfg}[1]{#1\pi G}
\newcommand{\teq}[1]{\hbox{$#1$}}

\newcommand{\Lm}{\mathcal{L}}
\newcommand{\Em}{\mathcal{E}}
\newcommand{\Um}{\mathcal{U}}

\newcommand{\nline}{\nonumber \\ }

\newcommand{\dV}{{\partial V}}
\newcommand{\SdV}{S_\dV}

\newcommand{\bint}{\sqrt{-g} \ n_\beta \frac{1}{g}\partial_\alpha(gg^{\alpha\beta})}

\begin{document}

\title{Statistical mechanical space-time decomposition as a key to a (non-perturbative) quantization prescription for gravity}

\author{P. A. Mandrin$^{1}$}

\affil{$^1$Independent Researcher, Siebnen, Switzerland}

\email{mandrin@hispeed.ch}

\begin{abstract}
By assuming gravity and matter to be subject to a joint statistical mechanical concept (JSMC) and interpreting Rindler horizon sections as open thermodynamic systems, one arrives at a specific new form of non-perturbative Lorentzian path integral quantisation in a compact space-time region, with well-determined gravitational measure, adequate fixing of boundary conditions and causal geometries. JSMC implies a space-time decomposition, leading to a sum over configurations in line with path integral methods. In these developments, we carefully distinguish the concepts of Boltzmann's statistical mechanics and the path integral.
\end{abstract}

\section{\label{sec:intro}Introduction}
We present here new developments in the scope of the joint statistical mechanical concept (JSMC) for gravity and matter, which was first proposed in~\cite{Mandrin_2023}. JSMC is the result of a theory-based attempt to mitigate uncertainties on how to treat gravity in a quantum mechanical environment. Typical uncertainties are, for instance: quantum versus non-quantum approach for gravity, quantisation method, quantum variables, quantum structure of space-time, matter coupling. This present effort is motivated in part by the current impossibility to resolve the uncertainties experimentally.

The main strategy of JSMC is to improve our understanding on the role played by matter in the context of gravity and then to draw conclusions for the quantisation of gravity. There are indications that common thermodynamic properties of gravity and matter could be of great help in understanding the role of matter. On one hand, the path integral quantisation of matter has an appealing statistical mechanics interpretation. On the other hand, gravity possesses a thermodynamic interpretation via the entropy and temperature of a Rindler horizon. The thermodynamic interpretation of gravity and the statistical mechanics interpretation of quantum matter can be unified to a joint statistical mechanics concept. In this context, the role of matter is clarified and a well determined path integral quantisation procedure for gravity can be found. We shall first review and elaborate in more detail how this unification naturally arises from fundamental thermodynamic reasoning. One of the main aspects is that finite sections of Rindler horizons must be treated as open thermodynamic systems involving grand canonical ensembles. In a second step, we show important implications of JSMC on the path integral formulation of gravity. This naturally involves a space-time decomposition method. One consequence is that the path integration is performed over configurations with well-defined causal structure, thus leading to a regularisation of the path integral. As a comparison, the benefits of this property have been discussed for Lorentzian causal dynamical triangulation~\cite{cdt, cdt4}, and the regularisation effect also has been investigated recently for Lorentzian simplicial path integral~\cite{Borissova_et_al}. In the developments of JSMC, the comparison between the Boltzmann-type partition function concept and the slightly different Feynman path integral concept is also discussed. In the present article, we shall restrict the underlying gravitational theory to general relativity (GR).

JSMC is tightly related to Hawking's path integral concept~\cite{Gibbons_Hawking} and is, at the same time, different, mainly due to the interpretation of finite horizon sections, the unified statistical mechanics concept, the measure obtained via low gravity considerations and a starting new effort for the statistical understanding of the path integral.
%

\section{\label{sec:arguments}Arguments for a joint statistical mechanics concept}

As is well-known, gravity has a thermodynamic interpretation. For any null surface, one can find Rindler coordinates such that the surface is transformed into a horizon with an entropy $S$ and emitting radiation at a temperature $T$, and the thermal heat across the horizon is related to the gravitational field equations~\cite{Unruh, Jacobson, Padmanabhan_2014, Jacobson_2016}. A Rindler horizon is equivalent to a black hole horizon~\cite{Hawking, Bekenstein, Wald, Wald_2001}. The temperature and entropy are connected via the ``Einstein'' surface term in Rindler coordinates \teq{\tilde{t},\tilde{x}^j}~\cite{Majhi_Padmanabhan, Padmanabhan_2012, Padmanabhan_2014},

\begin{equation}
\label{eq:A_sur_P}
\mathcal{A}_{sur} = - \frac{1}{16 \pi G} \int d^3 \tilde{x} \ \sqrt{-g} \ n_\beta \frac{1}{g}\partial_\alpha(gg^{\alpha\beta}) = \tilde{\varepsilon} \int d\tilde{t} \ TS,
\end{equation}

\noindent which is the surface term with fixed metric on a 3-dimensional Rindler horizon, with \teq{n_\beta=\delta^0_\beta} (\teq{c=1}), and \teq{\tilde{\varepsilon}=\pm 1}. Fixing \teq{\tilde{\varepsilon}=+1} may produce negative temperatures in some cases, this is equivalent to time reversal or entropy decrease while $\tilde{t}$ increases. If one wishes to keep the temperature always positive, one can achieve this by carefully choosing the sign~$\tilde{\varepsilon}$. The Rindler coordinates are obtained by boost transforming (local) inertial coordinates \teq{t_i,x_i^j}. In Rindler coordinates, (\ref{eq:A_sur_P}) is equal to the often used boundary term expression by York, Gibbons and Hawking~\cite{York, Gibbons_Hawking}. Although we restrict the underlying gravitational theory to GR, the concept of temperature (and thus (\ref{eq:A_sur_P})) can be adapted to more general theories of gravity~\cite{Kubiznak_Liska}.

\paragraph*{}
There are good reasons to consider finite sections of Rindler horizons rather than a full Rindler horizon. One reason is that a full Rindler horizon would have infinite entropy and is therefore ill-defined. Another reason is that the gravitational field should account for space-time variability (at least as an effective theory). And finally, a good unifying concept should, in the flat space-time limit, allow for the emergence of the usual path integral with space-time dependent matter fields. While a Rindler horizon is interpreted as a microcanonical ensemble, a finite section of it is an open thermodynamic system with respect to ``gravitational deformations'' and matter and should therefore be interpreted as a grand canonical ensemble~\cite{Mandrin_2023}. According to Boltzmann's statistical mechanics for a conventional gas at thermodynamic equilibrium, a grand canonical ensemble must have its grand canonical potential

\begin{equation}
\label{Phi_grand}
\Phi = \Um - TS - \mu N
\end{equation}

\noindent minimised, where $\Um$ is the average energy of the system, $\mu$ is the chemical potential and $N$ is the particle number. Since, for a microcanonical ensemble, $\Um$ would have to be minimised at equilibrium, and since, in GR, the energy is certainly not the wright quantity for a general minimisation principle, the analogue of $\Um$ must clearly be something completely different. For a large system (in terms of degrees of freedom), the gravitational equilibrium state must be the geometric configuration with the highest probabillity to occur, i.e. it must be a stationary point, and the only known such candidate is the saddle point of Lagrangian theory (besides the horizon area or entropy). Therefore, the analogue of $\Um$ must be the action (times a constant). This provides a good potential matching with \teq{TS} being a surface term of the action since completing the domain of integration to a full boundary provides the boundary term $S_{\dV}$ of the gravitational action, i.e. the gravitational analogue of the free energy \teq{F=\Um - TS} leads us to the Einstein-Hilbert action 

\begin{equation}
\label{eq:SH_SE}
S_H = S_E - \SdV, \qquad \SdV = - \frac{1}{16 \pi G} \int_\dV d^3 \tilde{x} \ \bint,
\end{equation}

\noindent where $S_E$ is the ``Einstein action'' which does not contain any second derivative of the metric, and the quantity to hold fixed on the boundary is according to~\cite{York} (metric or a certain derivative expression of it, depending on the chosen bulk action expression). Of course, we can go to non-Rindler coordinates $x^\alpha$ (with regular non-null boundaries) to compute \teq{S_H+\SdV} and evaluate the saddle point. $\SdV$ (like $S_E$) is frame-dependent, so that $\SdV$ in coordinates $x^\alpha$ does not normally correspond to \teq{TS}, unless one imposes the ``temperature correspondence relation''

\begin{eqnarray}
\sum_k \tilde{\epsilon}_k \int_{H_k} d\tilde{t} \ T_kS_k & = & -\frac{1}{16 \pi G} \int_\dV d^3 \tilde{x} \ \bint \nline
\label{eq:T_from_Sdv}
& \overset{!}{=} & -\frac{1}{16 \pi G} \int_\dV d^3x \ \bint,
\end{eqnarray}

\noindent where $H_k$ is the $k$th horizon constituent of the pure-horizon boundary in (local) Rindler coordinates, and horizon-specific quantities are designated with an index $k$. Indeed, (\ref{eq:T_from_Sdv}) is the answer to what temperature values represent thermodynamically the action with boundary term (in any frame). 

\paragraph*{}
We may now easily answer the question: If gravity leads us to consider grand canonical ensembles, what is the role of matter? The answer is found in (\ref{Phi_grand}). Since geometries in GR are fully determined by one tensorial field variable (metric or frame field) and the quantities in (\ref{eq:SH_SE}) are metric-dependent, the analogue of the third term of (\ref{Phi_grand}) can be nothing else than the matter term $S_m$ of the action. In $S_m$, since the globally conserved extensive quantities are the conserved charges of the gauge fields (appart from the energy-momentum), $N$ must be represented, at least in part, by the quantum numbers of the charges of matter ($N$ cannot be represented by the metric or its determinant since \teq{F=\Um - TS} would otherwise have to depend on $N$). Overall, one has to introduce a ``chemical potential'' term, at least for every conserved charge. As for the gravitational field, since no conserved charge can be given for a compact region $V$ in a specific frame, gravity cannot provide a ``chemical potential'' term as such. We therefore are led to attribute the path integral of matter to the ``chemical potential'' part of the grand canonical partition function. This is an appealing argument for JSMC since, without JSMC, we would have the difficulty of interpreting the ``chemical potential'' term resulting from Rindler horizon sections.

\section{\label{sec:stat2pi}Statistical mechanics versus path integral} 

In the regime where single particle effects are important, configurations aside of the maximum probability one gain in importance. With this in mind, we are looking for the analogue of the Boltzmann partition function

\begin{equation}
\label{PF_grand}
Z_{grand} = \sum_{i,N} \exp{[-\beta\Em_i+\beta\mu N]},
\end{equation}

\noindent where $\Em_i$ is the energy at the $i$th phase space point, $\beta=1/(k_B T)$ and $k_B$ is the Boltzmann constant. Since the analogue of $\Em_i$ is the action and the expression for the GR action is not bounded from above or below, (\ref{PF_grand}) cannot simply be transcribed to gravity. However, we may take advantage of the notion of path integral as used for matter fields. The path integral and the statistical mechanics partition function are considered to be tightly related to each other; this fact has been successfully used in various applications, e.g.~renormalisation theory. Indeed, in the ``low energy density regime'', we may neglect the contribution of $\Em_i$; this corresponds to flat space-time which is described by the path integral of quantum field theory (QFT),

\begin{equation}
\label{eq:path_qft}
Z|_{t_1\rightarrow t_2} = \int \ \prod_l \mathcal{D} \phi^{(l)} \ e^{i{S_m}|_{\phi^{(l)}(t_1)}^{\phi^{(l)}(t_2)}/\hbar},
\end{equation}

\noindent where the matter fields $\phi^{(l)}$ are running over species ($l$), and $\phi^{(l)}(t_1), \phi^{(l)}(t_2)$ are the  initial and final conditions. An important difference between (\ref{PF_grand}) with \teq{\Em_i=0} and (\ref{eq:path_qft}) is the imaginary factor~$i$ which turns the Boltzmann weighting factor into an oscillatory interference mechanism. Here, we shall not review the perspective which elegantly makes use of imaginary times and temperatures, but prefer another road adapted to the JSMC philosophy. As it should, the interference mechanism may be understood to have a weighting effect in a similar way as the Boltzmann factor has, with a maximum precisely at the stationary point. This is easily seen by summing exponentials evaluated with the functionals $\phi^{(l)}$ having nearby ``values'' (small ``difference'' \teq{\Delta \phi^{(l)}}),

\begin{equation}
\frac{1}{2}\bigg|e^{iS_m/\hbar}+e^{i(S_m+\Delta S_m)/\hbar}\bigg| = \sqrt{\frac{1}{2}+\frac{\cos{(\Delta S_m/\hbar)}}{2}},
\end{equation}

\noindent where $\Delta S_m$ is the difference of the action between the nearby ``values''. We see that the exponent of (\ref{PF_grand}) is equated with approximately \teq{-\Delta S_m^2/(8\hbar^2)} if \teq{\Delta S_m/\hbar \ll \pi/2}. That means, although the path integral acts with weightings in a similar way as the Boltzmann type partition function, it does not have the same dependence on the exponent. The weighting mechanism of the path integral must be considered as a statistical weighting for a theory with different constraints as compared to Boltzmann theory. We interprete the path integral as being the prescribed weighting mechnaism for quantisation, where the real quantity in the exponent, e.g. $S_m$, is the counterpart of the exponent in the partition function of Boltzmann statistical mechanics. For gravity with matter, (\ref{PF_grand}) and (\ref{Phi_grand}) tell us that we have to make the replacement \teq{S_m \rightarrow S_E + S_m} in the exponent of the path integral.

\paragraph*{}
Another aspect of (\ref{eq:path_qft}) which has to be managed is the space-time variability of the Lagrangian density. We also have such a variability for gravity. It appears that the gravitational field, its derivative and therefore the temperatures $T_k$ have a space-time variability due to (\ref{eq:T_from_Sdv}), i.e. the boundary $\dV$ of a compact region $V$ does not have constant temperature. The situation looks problematic at first sight: We are supposed to consider systems which are open with respect to the gravitational field, while $\dV$ does not have constant temperature. 

Again, this problem can be overcome by inspecting the path integral (\ref{eq:path_qft}) of QFT. As is well-known, this path integral can be computed by breaking up the path into tiny steps or, in other terms, by decomposing the space-time domain of integration into small local elements $\Delta V$ and integrating each subsystem over all field values at the corresponding space-time position. In thermodynamic language, we have to sum each subsystem ($\Delta V$) over all its thermodynamic states so that all subsystems together yield the full system ($V$). In (\ref{eq:path_qft}), the field configurations can be chosen to be smooth, but this still allows for arbitrary field configurations even though the value of the field function hardly varies as $\Delta V$ is made very small. 

The situation for the gravitational field is analogous. The elements $\Delta V$ should be such that $g_{\alpha\beta}$ and $g_{\alpha\beta,\gamma}$ (related to $T_k$) are approximately constant along each $k$-th horizon boundary component of $\Delta V$ in coordinates $\tilde{x}^\mu$ (see (\ref{eq:T_from_Sdv})). In order to achieve any desired accuracy by choosing small enough $\Delta V$-elements, we have to use smooth geometric configurations, i.e. charts. In that case only, an element with arbitrarily nearly constant temperature can be thought to be immerged in the thermal bath given by those neighbouring elements which share the same horizon surface and, indeed, very nearly the same temperature. In this way, a given geometric configuration of a $\Delta V$-element represents a grand canonical ensemble with fixed temperature $T_k$ for every horizon piece $H_k$ on the boundary of $\Delta V$.

\paragraph*{}
 We are now ready to generalise the correspondence between

\begin{equation}
\label{eq:path_qft_DV}
Z|_{\Delta V; t \rightarrow t+\Delta t} = \prod_l \sum_{\phi^{(l)}} \ e^{i
\int_{\Delta V}\Lm_m
|_{\phi^{(l)}(t)}^{\phi^{(l)}(t+\Delta t)}
/\hbar},
\end{equation}

\noindent and (\ref{PF_grand}) with \teq{\Em_i=0}. The generalisation to \teq{\Em_i\ne 0} is achieved by the replacement \teq{\Lm_m \rightarrow \Lm_E + \Lm_m} and by inserting one more sum in (\ref{eq:path_qft_DV}), by analogy to (\ref{PF_grand}), the sum is over all (functional) configurations of a gravitational variable $X$ (which is yet to be determined). Finally, to obtain the expression for the full region $V$, we must add up all elements $\Delta V_k$, sum for each $\Delta V_k$ over all values of $\phi^{(l)}$ and all values of the formal gravitational variable $X$. This yields the full path integral for gravity and matter (here as a formal expression; we ignore the boundary conditions for the moment):

\begin{equation}
\label{eq:path_g_firstansatz}
Z = \int \ \mathcal{D} X \prod_l \mathcal{D} \phi^{(l)} \ e^{i(S_E+S_m)/\hbar}.
\end{equation}

\section{\label{sec:measure}The measure}

In (\ref{eq:path_g_firstansatz}), it has been left open which gravitational variable to insert for $X$. A suggestion for this variable can be found from quantised linearised gravity together with the grand canonical ensemble interpretation of gravity~\cite{Mandrin_2023}. We first consider a single bosonic excitation mode $\phi$ (we restrict ourselves to bosonic matter to avoid the necessity to extend GR to account for spin). Then, for \teq{g_{\alpha\beta}\approx\eta_{\alpha\beta}}, we can use the standard canonical quantisation prescription to quantise the linearised Einstein field equations, this is equivalent to evaluating (\ref{eq:path_g_firstansatz}): 

\begin{equation}
\label{eq:EE_scalar_q}
\hat{G}_{\alpha\beta} + \Lambda \hat{g}_{\alpha\beta} = -\gfg{8} \hat{T}_{\alpha\beta},
\end{equation}

\noindent with cosmological constant $\Lambda$ and where the hat indicates that we have an operator. For \teq{\Lambda = 0}, using the ansatz

\begin{equation}
\label{eq:qg_sq_main}
\hat{g}_{\alpha\beta} = \eta_{\alpha\beta} + \hat{h}{}^I_{1\alpha}[\hat{\phi}] \ \eta_{IJ} \ \hat{h}{}^J_{1\beta}[\hat{\phi}] ( + \hat{h}{}^I_{2\alpha}[\hat{\phi}] \ \eta_{IJ} \ \hat{h}{}^J_{2\beta}[\hat{\phi}] + \ldots ) + \hat{h}_{L\alpha\beta}(x^\mu),
\end{equation}

\noindent where \teq{\hat{h}_{L\alpha\beta}(x^\mu)} is non-oscillatory and $\hat{\phi}$-independent, we find:

\begin{equation}
\label{eq:lin_h_phi_one_mode}
\hat{h}^I_{1\alpha} \sim i\sqrt{\pi G} \hat{\phi} \ ( \sim \hat{h}^I_{2\alpha}, \ldots ),
\end{equation}

\noindent i.e. $\hat{h}^I_{1\alpha}$, (\teq{\hat{h}^I_{2\alpha}, \ldots}) depend linearly on $\hat{\phi}$. Therefore, the canonical quantisation of $\phi$ induces an equal-time quantization prescription for linearised gravity describing the bosonic mode excitation:

\begin{eqnarray}
\label{eq:qh_sq_prescr1}
\left[\hat{h}{}^I_\alpha(x^i,t),\hat{h}{}^J_{\beta,0}(x'^i,t)\right] & \sim & - L_p^2 \delta(x^i-x'^i),\\
\label{eq:qh_sq_prescr2}
\left[\hat{h}{}^I_\alpha(x^i,t),\hat{h}{}^J_\beta(x'^i,t)\right] & = & \left[\hat{h}{}^I_{\alpha,0}(x^i,t),\hat{h}{}^J_{\beta,0}(x'^i,t)\right] = 0,
\end{eqnarray}

\noindent where $L_p$ is the Planck length and the $\delta$-function is restricted to 3 dimensions. (\ref{eq:qh_sq_prescr1}-\ref{eq:qh_sq_prescr2}) suggest that the correct variable to use for $X$ should be $h^I_\alpha$. We are not free in the choice of variable $X$ since imposing a modification of (\ref{eq:qh_sq_prescr1}-\ref{eq:qh_sq_prescr2}) as a quantisation prescription, e.g. inserting a power function of $\hat{h}^I_{\alpha}$, fails to hold simultaneously with (\ref{eq:qh_sq_prescr1}-\ref{eq:qh_sq_prescr2}), i.e. that would cause a contradiction to the standard model. Therefore, we have to make the replacement 

\begin{equation}
\label{eq:measure_h}
\mathcal{D} X \rightarrow \mathcal{D} h^I_\alpha
\end{equation}

\noindent in (\ref{eq:path_g_firstansatz}). The measure  \teq{\mathcal{D} h^I_\alpha} also must handle pure gravitational excitations $\hat{h}^I_{\alpha}$ which solve the equation (\ref{eq:EE_scalar_q}) with \teq{\hat{T}_{\alpha\beta}=0} (homogeneous solution). The same measure must hold for the non-perturbative theory as well since a different choice (e.g. a power function of $h^I_\alpha$) would contradict the standard model in the limit of linearised gravity.

\section{\label{sec:decomposition}Space-time decomposition}

There remain further open aspects, i.e. (i.) the boundary conditions in (\ref{eq:path_g_firstansatz}), (ii.) how space-time elements with pure horizon boundaries together with coordinates $\tilde{x}^\mu$ can be properly obtained, including the computation of the $T_k$, and (iii.) how the admitted geometric configurations for the path integral are impacted by the thermodynamic interpretation of gravity. 

\begin{figure}[h]
\begin{minipage}{16pc}
\includegraphics[width=16pc]{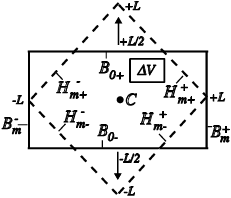}
\caption{\label{fig1}NNBP (solid lines) with boundary components $B_{0\pm},B^\pm_m$, and NBP (dashed lines) with null boundary components $H_{m\pm}^\pm$; the local inertial time coordinate is vertical, the local inertial spatial coordinate $x^m$ is horizontal.}
\end{minipage}\hspace{2pc}
\begin{minipage}{16pc}
\includegraphics[width=16pc]{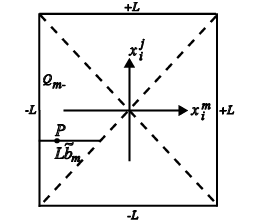}
\caption{\label{fig2}$(x^m_i,x^j_i)$-projection (local inertial coordinates) of a NBP with (projected) null-boundary intersection surfaces (dashed lines), exemplarily with negative-sided 2-surface $Q_{m-}$ nearest to the point $P$ and normal line from $Q_{m-}$ through $P$.}
\end{minipage} 
\end{figure}

These aspects must be addressed via the space-time decomposition into $\Delta V$-elements. Two different shapes are needed for every $\Delta V$-element. On one hand, the components of $\dV$ are usually non-null (in coordinates $x^\mu$). We then need to construct a brick-wall of $\Delta V$-elements, each with non-null boundaries, to approximate $V$ and $\dV$. For simplicity, we shall use rectangle (non-null boundary) polytope elements in terms of coordinates $x_i^\mu$ (NNBP elements). On the other hand, in order to exploit (\ref{eq:T_from_Sdv}), we also need to have null boundary polytopes, i.e. elements with pure null boundaries (in coordinates $x_i^\mu$) (NBP elements). The required coordinate transformations are performed in two steps, \teq{x^\mu\rightarrow x_i^\mu\rightarrow\tilde{x}^\mu}. Though being differently shaped, NNBP and NBP elements shall have the same 4-volume. NNBP are adapted to fit into $V$ and $\dV$, whereas NBP are adapted to thermodynamic language. The center point $C$ of each NNBP element shall coincide with the center point of the corresponding NBP element (in terms of $x_i^\mu$). If we set the center point $C$ of an element to be \teq{x_i^\mu=0}, the respective polytopes are defined by their corners which are, in terms of $x_i^\mu$-coordinates, \teq{(\pm L /2,\pm L,\pm L,\pm L)} for the NNBP element (all sign combinations, yielding half of a tesseract), while they are \teq{(\pm L,0,0,0)} and \teq{(0,\pm L,\pm L,\pm L)} for the NBP element (4d-analogue of the 3d-octohedron). In Figure~\ref{fig1}, a NNBP element and its corresponding NBP element are shown with two dimensions suppressed. 

When evaluating (\ref{eq:T_from_Sdv}), the transformation \teq{x_i^\mu\rightarrow\tilde{x}^\mu} must have been defined in such a way that the resulting NBP boundary components are pieces of horizons, and the transformation must be well-defined and smooth within the NBP. This can be achieved e.g. using the following transformation prescription, evaluated in a certain limit of ``tiny transitions'':

\begin{eqnarray}
\label{eq:tnbp}
t_i & = & f_m \bar{\bar{t}}^{m-} + (1-f_m) \bar{\bar{t}}^{m+}, \\
\label{eq:xnbd}
x^m_i & = & f_m \bar{\bar{x}}^{m-} + (1-f_m) \bar{\bar{x}}^{m+}, \\
\label{eq:f_fun}
f_m & = & \frac{1-\tanh{(\sigma\tilde{x}^m)}}{2}, \\
\label{eq:tbb}
\bar{\bar{t}}^{m\mp} & = & (L \pm \tilde{x}^m)[\kappa^\mp_{m\mp}\tilde{t}+\chi^\mp_m\{\sinh{(\kappa^\mp_{m\mp}\tilde{t})}-\kappa^\mp_{m\mp}\tilde{t}\}], \\
\label{eq:xbb}
\bar{\bar{x}}^{m\mp} & = & (\pm L + \tilde{x}^m)[1+\chi^\mp_m\{\cosh{(\kappa^\mp_{m\mp}\tilde{t})}-1\}] \mp L, \\
\label{eq:chi}
\chi^\mp_m & = & 1 - \tanh^\nu{[\rho \tilde{b}_m(L\pm\tilde{x}^m)\cosh{(\kappa^\mp_{m\mp}\tilde{t})}]},
\end{eqnarray}

\noindent where \teq{\kappa^\mp_{m\mp}=2\pi \tilde{\varepsilon}^\mp_{m\mp} T^\mp_{m\mp}}, the lower sign corresponds to the sign of $\tilde{t}$, the time reversal sign $\tilde{\varepsilon}^\mp_{m\mp}$ is as defined in (\ref{eq:A_sur_P}), no sum convention is understood for the hypersurface index $m$ which is defined so that the ``$m$-oriented Rindler vortex'' $Q_{m\pm}$ (2-surface \teq{x_i^m=\pm L}, \teq{t_i=0}) is nearest to the considered point $P$ with coordinates $x_i^\mu$, and the quantity \teq{L\tilde{b}_m\le L} is the projected inertial coordinate length of the normal line which goes from $Q_{m\pm}$ through $P$ up to the neighbouring null-boundary intersection surface. This is shown in a (\teq{x_i^m,x_i^j})-projection in Figure~\ref{fig2}, where \teq{j\ne 0,m}. The constants $L\sigma\gg 1$, $\rho>1$ and \teq{2<\nu\in\mathcal{N}} have been introduced to progressively approach full horizon coverage while maintaining a smooth transition near the intersections. For instance, one would start with \teq{\sigma=10/L}, \teq{\rho=2/L}, \teq{\nu=50} which yields $\approx80\%$ of horizon section coverage when \teq{\kappa^\mp_{m\mp}=1} and \teq{\tilde{b}_m=1}, and one would have to tune these constants to increase the horizon section coverage while preserving smooth transitions, thus approaching the limit of ``tiny transitions''. Moreover, in order to obtain finite temperatures $T_k$ for finite boundary term contributions, one would have to replace equation (\ref{eq:xbb}) by a regulating procedure, typically as follows:

\begin{eqnarray}
\label{eq:xnbp}
\bar{\bar{x}}^{m\mp} & = & \tilde{x}^m+(\pm L + \dot{x}^m)\chi^\mp_m\{\cosh{(\kappa^\mp_{m\mp}\tilde{t})}-1\}, \nline
\label{Rindler_reg}
\tilde{x}^m_i & = & \frac{\sqrt{\pi}[1-\textnormal{erf}(u)]+2ue^{-u^2}}{4 \sqrt{\kappa L}}; \qquad
u = \sqrt{\ln 2 - \ln \dot{x}^m}; \qquad
\textnormal{erf}(u) = \frac{2}{\sqrt{\pi}} \int e^{-u^2} du.
\end{eqnarray}

\noindent Each value \teq{T_k} may be computed using the NBP boundary term expression for the boundary component $H_k$ it belongs to, which in turn can be related with the NNBP boundary term expressions. Explicitly, we define the NNBP boundary components $B_{0\pm}$ to lie in the planes \teq{t_i = \pm L/2} $B_{0\pm}$ and $B_m^\pm$ in \teq{x_i^m = \pm L}, and we define the NBP boundary components  \teq{H_{m\pm}^-} to be delimited by the surfaces spanned between the 5 points \teq{(0,x_i^i)} with \teq{x_i^m=-L}, \teq{x_i^{i\ne m}=\pm L} and \teq{(\pm L,0,0,0)} (only one sign, matching the sign of the lower index of \teq{H_{m\pm}^-}), and \teq{H_{m\pm}^+} correspondingly (same but using \teq{x_i^m=+L} instead). The boundary components notations are shown in Figure~\ref{fig1}.

\noindent From (\ref{eq:A_sur_P}), we have \teq{\mathcal{A}_{sur} = -\tilde{\varepsilon}TLA_\perp}, where $A_\perp$ is the horizon area. By construction, evaluating the boundary term yields the same value for the NBP elements in coordinates $\tilde{x}^\mu$ as for the NNBP elements in coordinates $x^\mu$~\cite{Mandrin_2023}. We thus write for the boundary term contributions (taking into account the shape of the NBP):

\begin{eqnarray}
\label{eq:SdN}
S^\pm_{Hm+} & = & +\frac{1}{3} L^3 \tilde{\varepsilon}^\pm_{m+} T^\pm_{m+} = \frac{1}{2} (S^\pm_{Bm} + \frac{1}{3} S_{B0+}), \\
\label{eq:SdM}
S^\pm_{Hm-} & = & -\frac{1}{3} L^3 \tilde{\varepsilon}^\pm_{m-} T^\pm_{m-} = \frac{1}{2} (S^\pm_{Bm} + \frac{1}{3} S_{B0-}), \\
\label{eq:SdBi}
S^\pm_{Bm} & = & \pm \frac{1}{4\pi G} L^3 \sqrt{-g} [g^{i\mu}_{\ ,\mu(m\pm)} + g^{\rho\sigma}g^{m\mu}g_{\rho\sigma,\mu(m\pm)}], \\
\label{eq:SdB0}
S_{B0\pm} & = & \pm \frac{1}{2\pi G} L^3 \sqrt{-g} [g^{0\mu}_{\ ,\mu(0\pm)} + g^{\rho\sigma}g^{0\mu}g_{\rho\sigma,\mu(0\pm)}],
\end{eqnarray}

\noindent where the non-differentiated metric is the center metric (constrained by 4 gauge conditions) and the differentiated metric (constrained by \teq{4\times4} gauge conditions) approximately refers to the points \teq{x_i^{\beta}=\pm L \delta^\beta_\alpha} as specified by the index suffix in parentheses~``\teq{(\alpha\pm)}''. The equations (\ref{eq:SdN},\ref{eq:SdM}) are not independent but effectively imply four constraints for the \teq{T^\mp_{m\mp}}:

\begin{equation}
\label{eq:SdN_constr}
\tilde{\varepsilon}^+_{m\mp}T^+_{m\mp}-\tilde{\varepsilon}^-_{m\mp}T^-_{m\mp} = \tilde{\varepsilon}^+_{m\mp}T^+_{j\mp}-\tilde{\varepsilon}^-_{m\mp}T^-_{j\mp}.
\end{equation}

\noindent for any \teq{m,j=1,2,3}. (\ref{eq:SdN},\ref{eq:SdM}) follow from (\ref{eq:T_from_Sdv}) together with further constraints obtained by the fact that every null section is shared by two adjacent $\Delta V$-elements. Namely, the value $T^\mp_{m+}$ of one element is equal to the value $T^\pm_{m-}$ of the other element (the radiation into a future horizon is interpreted as the radiation out of a past horizon). Many $\Delta V$-elements (NNBP or NBP) can be lined up so that they are tightly glued to each other and cover the whole region $V$. This part answers the question of how to obtain the coordinates for pure horizon boundaries for every $\Delta V$-element and how to obtain the ``$T_k$''.

\paragraph*{}
We shall now address the question, which type of boundary conditions are required for the evaluation of the path integral (\ref{eq:path_g_firstansatz}). Indeed, all boundary components of the NBP or NNBP elements cancel out pairwise except for those along the outer boundary $\dV$. We therefore have $\dV$ at our disposal to impose conditions for $e^I_{\alpha}$ and $\phi^{(l)}$. Merely imposing initial and final conditions as in standard QFT would be the minimum boundary condition allowing to recover (\ref{eq:path_qft}) as a special case, but for compact $V$, we also may have a time-like boundary with specific reflection properties which are important for excitations propagating towards it. The same applies to QFT inside a box. The situation is already clear when considering the saddle point configuration, where time-like boundary conditions would be required in realistic situations to distinguish different reflection behaviours. We therefore adopt the imposition of full boundary conditions for $e^I_{\alpha}$ and $\phi^{(l)}$ in the general case, which we shall denote with the notation \teq{[\ldots]_{\partial V}}:

\begin{equation}
\label{eq:path_g_def}
Z\big|_{\partial V} = \int \mathcal{D} h^I_{\alpha} \prod_l \mathcal{D} \phi^{(l)} \ e^{i[S_H+S_{\dV}+S_m]_{\partial V}/\hbar}.
\end{equation}

\paragraph*{}
Finally, the thermodynamic interpretation of gravity has an impact on what geometric configurations are allowed to be summed over in (\ref{eq:path_g_def}). By construction and from JSMC, each NBP must have a well-defined causal structure in order for a transformation like (\ref{eq:tnbp}--\ref{eq:chi}) to be well-defined. Since every geometric configuration must be smooth and is the join of all $\Delta V$-elements, it follows that every geometric configuration must have a well-defined globally causal structure as well. This requirement restricts the allowed geometries by excluding changes in the spatial topology.
%

\section{\label{sec:conclusion}Conclusions}

The goal to theoretically mitigate the uncertainty on how to quantise gravity (and whether to quantise it) has led us to propose a joint statistical mechanical interpretation of gravity and matter, the argument has been strengthened in the first part of this article. The thermodynamics of Rindler horizons and the path integral quantisation of QFT have then been brought together and the immediate implications of the joint concept on the development of a quantum theory have been considered in more detail. The joint concept basically leads to a path integral expression. But there is much more information to gain from it: One finds that (i.) the gravitational measure is \teq{\mathcal{D} h^I_{\alpha}}, where the metric is linearly related to the bilinear form of $h^I_{\alpha}$, (ii.) the boundary condition for the path integral is to fix $e^I_{\alpha}$ and the matter fields on $\dV$, and (iii.) the sum over geometries is restricted to well-defined globally causal structures.
These findings have been obtained by making use of a space-time decomposition -- as implied by the correspondence between thermodynamics and path integrals. The space-time decomposition also allows to explicitly compute the temperature values of every horizon section on the (adapted) Rindler boundaries of space-time elements.


\end{document}